\documentclass[epj]{webofc}
\usepackage[utf8]{inputenc}
\usepackage[varg]{txfonts}   
\usepackage{booktabs}
\usepackage{xcolor}
\definecolor{darkred}{rgb}{0.4,0.0,0.0}
\definecolor{darkgreen}{rgb}{0.0,0.4,0.0}
\definecolor{darkblue}{rgb}{0.0,0.0,0.4}
\usepackage[bookmarks,linktocpage,colorlinks,
    linkcolor = darkred,
    urlcolor  = darkblue,
    citecolor = darkgreen]{hyperref}
\usepackage{subfigure}
\wocname{EPJ Web of Conferences}
\woctitle{Lattice2017}
\newcommand{\eq}{\begin{equation}}
\newcommand{\en}{\end{equation}}
\newcommand{\eqa}{\begin{eqnarray}}
\newcommand{\ena}{\end{eqnarray}}

\newcommand{\SU}{\mathrm{SU}}
\newcommand{\xiratio}{\xi/\xi_{2nd}}

\begin{document}
%
\selectlanguage{english}
\title{%
The $\xiratio$ ratio as a test for Effective Polyakov Loop Actions
}
\author{%
\firstname{Michele} \lastname{Caselle}\inst{1,2}\fnsep\thanks{Speaker, \email{caselle@to.infn.it}} \and
\firstname{Alessandro} \lastname{Nada}\inst{1} 
}
\institute{%
Department of Physics, University of Turin \& INFN,  V. Giuria 1 10125 Torino (Italy)
\and
Arnold-Regge Center, University of Turin, V. Giuria 1 10125 Torino (Italy)
}
\abstract{%
Effective Polyakov line actions are a powerful tool to study the finite temperature behaviour of lattice gauge theories. They are much simpler to simulate than the original (3+1) dimensional LGTs and are affected by a milder sign problem. However it is not clear to which extent they really capture the rich spectrum of the original theories, a feature which is instead of great importance if one aims to address the sign problem. We propose here a simple way to address this issue based on the so called second moment correlation length $\xi_{2nd}$. The ratio $\xiratio$ between the exponential correlation length and the second moment one is equal to 1 if only a single mass is present in the spectrum, and becomes larger and larger as the complexity of the spectrum increases. Since both $\xi_{exp}$ and $\xi_{2nd}$ are easy to measure on the lattice, this is an economic and effective way to keep track of the spectrum of the theory. In this respect we show using both numerical simulation and effective string calculations that this ratio increases dramatically as the temperature decreases. This non-trivial behaviour should be reproduced by the Polyakov loop effective action. 
}
\maketitle

\section{Introduction}

In the past few years there has been a growing interest on the effective Polyakov loop (EPL) models for the description of QCD at finite temperature and finite chemical potential
\cite{Gattringer:2011gq,Mercado:2012ue,Greensite:2012xv,Greensite:2012dy,Greensite:2013yd,Greensite:2013bya,
Greensite:2014isa,Hollwieser:2016hne,Hollwieser:2016yjz,Langelage:2010yr,Fromm:2011qi,Bergner:2013qaa,Bergner:2015rza,Dittmann:2003qt,Heinzl:2005xv,Wozar:2006fi,
Wozar:2007tz,Billo:1996wv,Aarts:2011zn,Scior:2014zga,Bahrampour:2016qgw}.
In EPL models the original theory regularized on the lattice is mapped to a three-dimensional, center-symmetric, effective Polyakov loop spin model obtained by integration over the gauge and matter degrees of freedom. 
An exact integration of time-like degrees of freedom is too difficult, but from strong-coupling expansions one can infer a few important features of the action of such effective models:
\begin{itemize}
 \item first, it should be non-local, since, as the order of the strong-coupling expansion increases, far apart Polyakov loops are involved in the interaction;
 \item second, it should involve higher representations of the Polyakov loop;
 \item and third, it should contain multi-spin interactions.
\end{itemize}

Since in most of the existing proposals multi-spin interactions are neglected, with the assumption that they can be taken into account by a suitable tuning of two-spin interactions, the action takes the form
\begin{equation}
S_{\text{eff}}=\sum_{p}\sum_{|{\bf r}|\geq 1}\sum_{|{\bf x}-{\bf y}|=\bf r} \lambda_{p,{\bf r}} \, \chi_p({\bf x}) \, \chi_p({\bf y})
\label{linansatz}
\end{equation}
where $\chi_p (\bf x)$ is the character in the $p$ representation of the loop in the spatial site $\bf x$ and $\lambda_{p,{\bf r}}$ is the coupling between effective spins at distance ${\bf r}$.

Several strategies have been devised in recent years to address the proble of determininig the infinitely many interaction terms $\lambda_{p,{\bf r}}$ between the degrees of freedom in the new three-dimensional model.
The general idea of EPL proposals is to keep the number of free parameters in Eq.~\eqref{linansatz} as small as possible, trying to summarize in a few relevant couplings the complexity of the original gauge theory.
It is thus important to develop new tools that are able to test the ability of such proposals to capture relevant features of the original theory.
In a recent paper~\cite{Caselle:2017xrw} we suggested the use of the ratio between the exponential correlation length $\xi$ and the second moment correlation length $\xi_{2nd}$ as a tool to test EPL actions. 
This ratio is well defined for any model and it can be easily computed in Monte~Carlo simulations on the lattice.

This proposal has been inspired by the observation that any EPL model proposal must face two non-trivial requirements, which are, indeed, two faces of the same coin.

The first requirement is that Polyakov loop correlators extracted from EPL models should display the so-called ``L\"uscher term''~\cite{Luscher:1980fr}, i.e. a $1/R$ correction in the static quark-antiquark potential.
Such term is present in the confining phase of the original theory, and has been detected and precisely measured on the lattice in $\SU(N)$ gauge theories, both using Wilson loops~\cite{Necco:2001xg} and Polyakov loop correlators \cite{Luscher:2004ib,Caselle:2004er,Athenodorou:2010cs,Athenodorou:2011rx}; thus, EPL actions should reproduce the same behaviour. 
This is indeed a very non-trivial requirement, since such a term is typical of extended gauge invariant observables and, in general, spin models with short-distance interactions do not possess such a behaviour.

The second issue is that the original lattice gauge theory (LGT) is characterized by a rich spectrum of excitations, with an ``Hagedorn'' type of dependence on the energy, which, again, is typical of gauge theories and not easily to reproducible with a spin model.

These two features are deeply related, since it is precisely the accumulation of an infinite number of excitations that leads to the $1/R$ correction in the interquark potential. 
In the framework of the effective string description of LGTs, this relation can be shown explicitly, using the open-closed string duality or, equivalently, with a modular transformation of the effective string result for the interquark potential. 
We shall address this issue in Section~\ref{sec:est} of this contribution.

The $\xi/\xi_{2nd}$ ratio represents a simple and easy way to keep track of the spectrum of a statistical model: one can use this easily computable quantity to understand if the spectrum, both in the original theory and in the effective model, is dominated by a single mass or contains several masses in competition among them. 
This information could help in the selection of the terms to be included in the effective action and, possibly, even in the fine-tuning of the couplings obtained with existing approaches.
As an example, we will discuss the case of the $(3+1)$-dimensional $\SU(2)$ Yang-Mills theory and compare the results with the simplest possible EPL model for this theory, namely the nearest-neighbor Ising model in 3 dimensions.
We shall see that the Ising model is a good effective description of the original $\SU(2)$ theory only in the vicinity of the deconfinement transition, but becomes worse and worse as the temperature decreases.
Another, physically relevant example of this strategy was recently discussed in~\cite{Greensite:2017qfl}, where the authors studied a Polyakov Loop effective action derived using the relative weights method on a $\SU(3)$ gauge theory with dynamical staggered fermions of mass 695 MeV. 
The main choice the authors made in constructing this EPL mdoel was the inclusion in the action of large distance couplings between Polyakov loops. 
The authors found a value $\xiratio=1.27(3)$ compatible with a rich string-like spectrum, thus confirming the reliability of their choice.  

\section{The relation between \texorpdfstring{$\xi$}{} and \texorpdfstring{$\xi_{2nd}$}{} in spin models}
\label{sec:xi2nd_ising}

In a $d$-dimensional spin model the exponential correlation length $\xi$ describes the long distance behavior of the connected two point function and is defined as
\begin{equation}
\label{exp_corr_length}
 \frac{1}{\xi}=-\lim_{|\vec{n}|\to\infty}\frac{1}{|\vec{n}|} \log\langle s_{\vec{0}} s_{\vec{n}}\rangle_c
\end{equation}
where $s_{\vec{x}}$ denotes the spin $s$ in the position $\vec{x}$ and the connected correlator is defined as
\begin{equation}
\langle s_{\vec{m}} s_{\vec{n}} \rangle_c = 
\langle s_{\vec{m}} s_{\vec{n}} \rangle -  \langle s_{\vec{m}}  \rangle ^2.
\end{equation}
The square of the second moment correlation length $\xi_{2nd}$ is defined as:
\begin{equation}
\xi_{2nd}^2 = \frac{\mu_2}{2 d \mu_0} \;,
\label{xi2nd}
\end{equation}
where 
\begin{equation}
\mu_0 = \lim_{L \rightarrow \infty}\; \frac{1}{V} \; \sum_{{\vec{m}},{\vec{n}}} \;
\langle s_{\vec{m}} s_{\vec{n}} \rangle_c \hskip 1cm
\mu_2 = \lim_{L \rightarrow \infty}\; \frac{1}{V} \;  \sum_{{\vec{m}},{\vec{n}}} \; |\vec{m} - \vec{n}|^2 \langle s_{\vec{m}} \, s_{\vec{n}} \rangle_c \;,
\end{equation}
where $V= L^d$ is the lattice volume and the sum on $\vec{m}$ is on the $d$ indices $m_0,m_1,...,m_{d-1}$.

It can be shown (see~\cite{Caselle:2017xrw} for a detailed derivation) that if we assume a multiple exponential decay for the zero momentum correlator $G(\tau)$:
\begin{equation}
\langle S_0 \; S_\tau  
\rangle_c \propto \sum_i \; c_i \; \exp(-|\tau|/\xi_i) \;\; , 
\label{3mass}
\end{equation}
then
\begin{equation}
\label{e24}
 \xi_{2nd}^2 = \frac{ \sum_i c_i \xi_i^3}{\sum_i c_i \xi_i} \;,
\end{equation}
which is equal to $\xi^2$ if only one state contributes. 
It is thus clear that we can use the $\xi/\xi_{2nd}$ to have some insight on the spectrum of the theory and on the amplitude $c_i$ of these states.

In the following we examine a set of results for the Ising universality class that are reported in Table~\ref{tab:xi_ising}: this class of models has not only been studied in great detail and with high accuracy, but it is also the most relevant one for the study in the $\SU(2)$ pure gauge theory that we will discuss in Section~\ref{sec:su2results}.
The $d=2$ results are obtained from the exact solution of the two-dimensional Ising model while the $d=3$ results are obtained from $\epsilon$ expansion calculations, Monte~Carlo simulations or strong-coupling expansions.  
For a review of these and other results see for instance Ref.~\cite{Pelissetto:2000ek}.

\begin{table}[!htb]
\centering
\caption{Values of the $\xiratio$ ratio for an Ising spin system in three different conditions: in the high-temperature symmetric phase, in the low-temperature broken symmetry phase and along the critical isotherm. 
It is important to notice that in $d=3$ there is a general agreement among results obtained with very different approaches, ranging from Monte~Carlo simulations to strong-coupling expansions. 
\label{tab:xi_ising}}
\begin{tabular}{lccl}
\toprule
 & $d$ & $\xiratio$ & Method\\
\midrule
High-$T$ phase & $2$ & $1.00040...$  & \\
               & $3$ & $1.00016(2)$  & Strong-coupling + $\epsilon$-exp.~\cite{Campostrini:1999at}\\
               & $3$ & $1.00021(3)$  & Perturbative calc.~\cite{Campostrini:1997sn}\\
               & $3$ & $1.000200(3)$ & Strong-coupling~\mbox{\cite{Campostrini:1999at}}\\
\midrule
Low-$T$ phase & $2$ & $1.58188...$ & \\
              & $3$ & $1.031(6)$   & MC simulations~\cite{Caselle:1999tm}\\
              & $3$ & $1.032(4)$   & Strong-coupling~\mbox{\cite{Campostrini:1999at}}\\
\midrule
Critical isotherm & $2$ & $1.07868...$ & \\
($t = 0$,  $|H|\not= 0$) & $3$ & $1.024(4)$   & Strong-coupling + $\epsilon$ exp.~\mbox{\cite{Campostrini:1999at}}\\
\bottomrule
\end{tabular}
\end{table}

In the high-$T$ symmetric phase, where the spectrum is composed by a single massive state, we would expect that $\xi/\xi_{2nd}=1$: the small but not negligible difference from $1$ can be better understood looking at the Fourier transform of $G(\tau)$.
Besides isolated poles, which correspond to isolated exponentials in $G(\tau)$, we also have cuts above the pair production threshold at momentum $p$ equal twice the lowest mass. 
In the original correlator these cuts can be thought of as the coalescence of infinitely nearby exponentials\footnote{As a matter of fact on a finite lattice, this is their correct description, since the transfer matrix has only a finite number of eigenvalues.} and as such they also contribute to the ratio $\xi/\xi_{2nd}$.

In $d=3$, both in the low-$T$ broken symmetry phase and on the critical isotherm curve ($T=T_c$, $H\not=0$), the $\xi/\xi_{2nd}$ ratio is definitely larger: indeed, besides the cuts discussed above, we also have one (or more) isolated bound states which contribute to the spectrum. 
This is the case for instance of the 3-dimensional Ising model for $T<T_c$, for which an infinite tower of bound states exists \cite{Caselle:2001im}: in particular, the lowest of such states takes the value $m_{bound}=1.83(3) \, m_{ph}$~\cite{Caselle:1999tm} and is thus located below the two-particle threshold. 

The $d=2$, $T=T_c$, $H\not=0$ case is of particular interest: thanks to the exact solution of Zamolodchikov~\cite{Zamolodchikov:1989fp} we know that there are three particles in the spectrum below the two-particle threshold and accordingly the difference $\frac{\xi}{\xi_{2nd}}-1$ is about three times larger than in the $d=3$, $T<T_c$ case which, as we have seen, has only one state below threshold.

Finally, it is very instructive to look at the $d=2$, low-$T$ case, in which the Fourier transform of the correlators starts with a cut. 
Following the analysis of McCoy and Wu~\cite{McCoy:1978ta} (and more recently of Fonseca and Zamolodchikov~\cite{Fonseca:2001dc}) we know that we may interpret the spectrum of this model as the coalescence of an infinite number of states. 
Accordingly, a much larger value of the ratio $\xi/\xi_{2nd}$ is found, with a difference from $1$ which is one order of magnitude larger than the value in presence of an isolate bound state and three orders of magnitude larger than the $T>T_c$ case.

\section{Numerical results for the (\texorpdfstring{$D=3+1$}{}) \texorpdfstring{$\SU(2)$}{} lattice gauge theory}
\label{sec:su2results}

As a test of our proposal we studied the $\xiratio$ ratio in the $(D=3+1)$ $\SU(2)$ gauge theory. 
Results are reported in Table~\ref{tab:xi_xi2nd}. Further details on the simulation setting and on the data analysis can be found in~\cite{Caselle:2017xrw}.

\begin{table}[!ht]
\centering
\caption{Results for the $\xi$, $\xi_{2nd}$ and their ratio $\frac{\xi}{\xi_{2nd}}$ in the confined phase of the $(3+1)$ dimensional $\SU(2)$ gauge theory.
\label{tab:xi_xi2nd}}
\begin{tabular}{cccccc}
\toprule
$\beta$ & $T/T_c$ & $ N_s^3 \times N_t$& $\xi/a$ & $\xi_{2nd}/a$ & $\xi/\xi_{2nd}$\\
\midrule
$2.27$ & $0.59$ &  $32^3 \times 6$  & $1.31(2)$ & $0.887(8)$  & $1.48(3)$ \\
$2.33$ & $0.71$ & $32^3 \times 6$  & $2.31(4)$ & $1.842(15)$ & $1.25(2)$ \\
$2.30$ & $0.78$ &  $32^3 \times 5$  & $2.56(2)$ & $2.22(1)$   & $1.153(11)$ \\
$2.357$ & $0.78$ & $32^3 \times 6$  & $3.08(4)$ & $2.67(2)$   & $1.151(16)$ \\
$2.25$ & $0.84$ & $32^3 \times 4$  & $3.05(6)$ & $2.74(4)$   & $1.11(3)$ \\
$2.40$ & $0.90$ & $32^3 \times 6$  & $6.9(2)$  & $6.6(3)$    & $1.04(6)$  \\
\bottomrule
\end{tabular}

\end{table}

\begin{figure}
\centering
\includegraphics[scale=1]{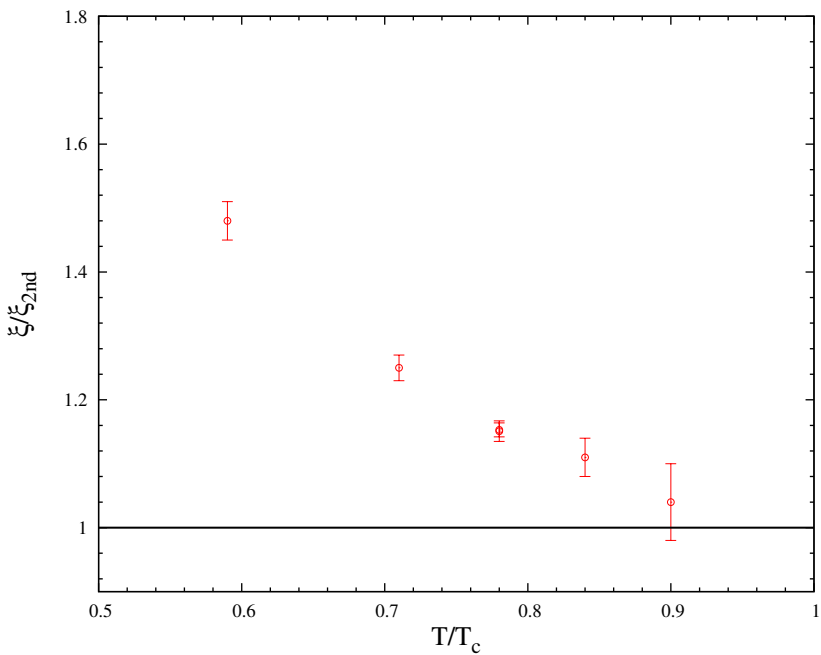}%
\caption{The $\xiratio$ ratio for different values of the temperature $T/T_c$ in the confining region.
\label{fig:xi2nd}
}
\end{figure}

The data show a clear dependence on $T/T_c$ and in particular a dramatic increase of $\xiratio$ as the temperature decreases. 
In order to test if there are other dependences in our results we realized the $T/T_c=0.78$ case with two different combinations of $\beta$ and $N_t$. 
We found the same value of $\xiratio$ even if the values of $\xi$ and $\xi_{2nd}$ were quite different in the two cases: this result makes us confident that scaling corrections are under control and that our results are tracing a true physical behaviour of the $\SU(2)$ model.

The simplest possible EPL model for the $\SU(2)$ lattice gauge theory discussed in the previous section is the $d=3$ Ising model, which corresponds to the case in which in Eq.~(\ref{linansatz}) we truncate the action to the nearest-neighbour term, choose only the fundamental representation and approximate the Polyakov loop with its sign. 
This is a very crude truncation, but in the $\SU(2)$ case, in the vicinity of the deconfinement transition it turns out to be a very good approximation. 
In fact, the Svetitsky-Yaffe conjecture~\cite{Svetitsky:1982gs} tells us that if a $(d+1)$ LGT with gauge group $G$ has a second order deconfinement transition and the $d-$dimensional spin model with a global symmetry group being the center of $G$ has a second order magnetization transition, then the two models belong to the same universality class. 
This is exactly the case of the $(D=3+1)$ $\SU(2)$ LGT and of the $d=3$ Ising model. 
Using symmetry arguments it is easy to see that the Polyakov loop (which is the order parameter of the deconfinement transition) is mapped by this identification into the spin of the Ising model (which is in fact the order parameter of the magnetization transition) and that the confining phase (the one which we studied in the previous section) is mapped into the high-$T$ symmetric phase of the Ising model.
From the discussion of Section~\ref{sec:xi2nd_ising} we thus expect the $\xiratio$ ratio to be very close to $1$ when the $\SU(2)$ model is close to the deconfinement transition, as this is the observed behaviour in the high-$T$ phase for the 3d Ising model (as reported in Table~\ref{tab:xi_ising}).
This is indeed the case for the highest value of $T/T_c$ that we simulated. 
However we see from the data that the ratio increases dramatically as $T/T_c$ decreases. 
This shows that the as $T/T_c$ decreases the Ising approximation becomes indeed too crude and more sophisticated EPL models are needed. 
We will see below that this increase in the $\xiratio$ is essentially due to the combination of two non-trivial features of the gauge theory spectrum: 
\begin{itemize}
\item
first, that as $T/T_c$ decreases the states of the spectrum coalesce toward the ground state, exactly as it happens in the $d=2$ Ising model below $T_c$;
\item
second, that the overlap constants $c_i$ increase exponentially with the energy of the states.
\end{itemize}
The nearest neighbour Ising model cannot mimic these two features and thus must be discarded as $T/T_c$ decreases. 
It is interesting to notice as a side remark that our analysis offers a nice and simple way to estimate the range of validity of the the Svetitsky-Yaffe conjecture which, within our range of precision holds for $T/T_c>0.9$.

\section{Effective String description of the interquark potential}
\label{sec:est}

A very useful tool to understand both the features of the spectrum mentioned above is the effective string description of the Polyakov loop correlators~\cite{Luscher:1980ac,Luscher:1980fr} which indeed predicts, as a consequence of the ``string'' nature of the color flux tube, a rich spectrum of excitations. 
In particular, it has been recently realized that the Nambu-Got\={o} action~\cite{Nambu:1974zg,Goto:1971ce} is a very good approximation of this effective string model~\cite{Luscher:2004ib,Aharony:2013ipa}. 
The Nambu-Got\={o} action is simple enough to be exactly solvable so that the spectrum can be studied explicitly: for example, the large distance expansion of the Polyakov loop correlator in $D$ space-time dimensions is \cite{Luscher:2004ib, Billo:2005iv}:

\begin{equation}
  \left\langle P(x)^{\ast}\kern-1pt P(y)\right\rangle = \sum_{n=0}^{\infty}\frac{N_t}{2r} w_n e^{-{E}_nr} \;
\end{equation}
where the weights $w_n$ can be obtained from the power expansion of the Dedekind function and the Energy levels $E_n$ can be written in terms of the temperature $T/T_c$ as follows (see \cite{Caselle:2017xrw} for a detailed derivation):
\begin{equation}
  {E}_n= \frac{2\pi T_c^2}{3 T} \left\{ 1 + 12 \, \frac{T^2}{T^2_c}\left[n-\frac{1}{12}\right] \right\}^{1/2}.
\label{eq:masses}
\end{equation}
This equation gives us a concrete realization of the two non-trivial features of the spectrum that we mentioned above:
\begin{itemize}
\item
first, the gap between the different states, {\sl decreases} as $T/T_c$ decreases and all the states tend to accumulate toward the lowest state.
\item
second, it can be shown that the amplitudes $w_n$ increase exponentially with $n$. 
\begin{equation}
w_n \sim \exp \{ \pi\sqrt{\frac{2(D-2)n}{3}} \} \;.
\label{eq26}
\end{equation}

\end{itemize}


The combination of these two effects drives the $\xi/\xi_{2nd}$ ratio to larger values as the temperature decreases, as represented in fig.~\ref{fig:xi2nd}. 
It is this behaviour which EPL actions should be able to mimic and in our opinion it represents a stringent test for existing proposals.

Notice, as a side remark, that it is exactly the accumulation of infinite massive excitations toward the lowest state which drives the ``L\"uscher'' $1/R$ term in the low-$T$ regime of the theory. 
It is unlikely that such a term could be obtained by any mechanism other than an accumulation of infinite poles, in a local 3d spin theory like the existing proposal for EPL actions. 
It is also interesting to observe that the values that we measure for the  $\xi/\xi_{2nd}$ ratio in the $\SU(2)$ lattice gauge theory are significantly larger than the single state expectation even for temperatures in the range $0.6<T/T_c<0.9$ where (see Eq.~(\ref{eq:masses})) there are no masses below the two particle threshold. 
This large value is due to the exponential increase of the weights of the excited states with their energy as shown in Eq.~(\ref{eq26}). 
This is a typical ``string-like'' effect and it is exactly this type of signatures that the EPL model should be able to reproduce.

\section{Concluding remarks}
\label{sec:conclusion}

It is interesting to notice that there is a natural implementation in effective Polyakov loop actions of the infinite tower of excited states discussed in Section~\ref{sec:est}: these are the traces of the Polyakov loop in representations higher than the fundamental one. 
These terms naturally appear in the strong-coupling expansion: they are subleading and are thus usually considered as negligible, but we expect that they should become more and more important as the temperature decreases. 
Indeed it was recently observed~\cite{Bergner:2015rza} that, as the temperature decreases, higher representation terms in the effective action become more and more important, and their contribution is not compensated by an increase in the distance of couplings in the fundamental representation.
At the same time, it is likely that EPL models with long range interactions with a power-like decrease of the coupling constants (as those recently proposed in refs.~\cite{Greensite:2013yd,Greensite:2013bya,Greensite:2014isa,Hollwieser:2016hne}) could be characterized by a much richer spectrum than standard nearest-neighbour models and they could represent another strategy to obtain larger values of the $\xiratio$ ratio. 
This was recently confirmed by a direct calculation in Ref.~\cite{Greensite:2017qfl}, where the authors studied the $\xiratio$ ratio in an EPL action with long range couplings, derived using the relative weights method on a $\SU(3)$ gauge theory with dynamical staggered fermions of mass 695 MeV. 
They found $\xiratio=1.27(3)$ which is indeed compatible with a rich string-like spectrum. 

\bibliography{csi2}

\end{document}